# Surface Plasmons and Field Electron Emission In Metal Nanostructures

N. Garcia[1] and Bai Ming[2]

1. Laboratorio de Física de Sistemas Pequeños y Nanotecnología, Consejo Superior de Investigaciones Científicas, Serrano 144, Madrid 28006 (Spain)
and Laboratório de Filmes Finos e Superfícies, Departamento de Física, Universidade Federal de Santa Catarina, Caixa Postal 476, 88.040-900, Florianópolis, SC, (Brazil)
2. Electromagnetics laboratory, School of Electronic Information Engineering, Beijing University of Aeronautics and Astronautics, Beijing, 100191 (China)

**Abstract**

In this paper we discuss the field enhancement due to surface plasmons resonances of metallic nanostructures, in particular nano spheres on top of a metal, and find maximum field enhancement of the order of $10^2$, intensities enhancement of the order of $10^4$. Naturally these fields can produce temporal fields of the order of 0.5V/Å that yield field emission of electrons. Although the fields enhancements we calculated are factor of 10 smaller than those reported in recent experiments, our results explain very well the experimental data. Very large atomic fields destabilize the system completely emitting ions, at least for static field, and produce electric breakdown. In any case, we prove that the data are striking and can solve problems in providing stabilized current of static fields for which future experiments should be done for obtaining pulsed beams of electrons.

Electrons are emitted from surfaces due to the photoelectron effect when the light energy that illuminates the surface is larger than its work function. But also more interesting with the development of pulsed lasers are the multiphoton processes[1,2]. However what is striking is when the illumination has huge laser intensity such that its field can be large enough to produce field electron emission from the surface directly. That is huge laser intensities can produce field electron emission from surfaces.

It is known since long time ago that when light illuminate a metallic surface having nanostructures comparable to the wavelength of the light and on resonant conditions, the field of the incident light can increase locally, on the surface, by large factors due to surface plasmon excitations [2-5]. This idea was used to explain the large increase in the Raman effect of molecules in surfaces [6]. Naturally this is a very interesting effect if one notices the additional physical problems related to the increase of the field. This is precisely what has been reported in a recent paper by Schertz et al. [7], who have observed how electrons are emitted from a structure consisting of an Au sphere on top of an Au surface isolated by a dielectric. They observed electron emission when the structure was illuminated by a pulsed laser at a given angle of incidence. These results are striking and can be of importance to stabilize field emission currents of previous applied static field modulated by pulsed laser illuminations.

We would like to discuss these data [7], of Au spheres on top of Au surfaces, and explain, from our results, what are the ideal conditions to have the field emission. We find that for the conditions of the experiments the field enhancement found $E_{res}/E_{inc}$ (the ratio of the electric field in resonance divided by the incident field) has a maximum of approximately $1.5 \times 10^2$ and not $10^3$ as claimed. However this is not a problem to explain the observed electron emission to agree with the experimental data. Note that the reported fields of 5V/Å [7] are atomic fields, which are capable of removing material, destructs the structure and produces breakdown for Au [8], at least as static fields. To describe our results, we proceed by calculating first the field enhancements due to excitation of plasmon resonances and then the field emission for the obtained fields. Finally we conclude the results.

To analyze the near field enhancement of the Au particle on an Au plate, a full wave simulation of plasmon excitation is required. Finite difference time domain (FDTD) method has been widely used [5,9] and proven to be an efficient tool to simulate the plasmons in nano-particle scenario [10,11]. The FDTD method numerically solves Maxwell's equations in discrete space by time steps. It is capable of tracing the dynamic evolution of the electromagnetic field to obtain the system response in time or in spectrum. Therefore to analyze the plasmon resonance, one can launch a broadband short pulse as excitation, record the electric field at the observation point in time and then obtain the spectrum response by Fourier transform.

The dynamic of plasmons is governed by the metallic properties in the studied frequency region, which is the dispersive permittivity $\varepsilon = \varepsilon(\omega)$ in

$$\nabla \times \bar{H} = \frac{\partial \bar{D}}{\partial t}, \quad \bar{D} = \varepsilon_0 \varepsilon(\omega) \bar{E}. \tag{1}$$

To describe the noble metal Au, a lossy Drude dispersive model is used: $\varepsilon(\omega) = \varepsilon_\infty - \frac{\omega_p}{\omega^2 + j2\omega\gamma}$, where high frequency response $\varepsilon_\infty = 12.18$, the plasma frequency $\omega_p = 1.447 \times 10^{16}$, and the damping coefficient $\gamma = 1.406 \times 10^{14}$. This model is more in mathematical sense and is well fitted with the measured dielectric constant of Au in near infrared range [12,13].

The Maxwell's equations with dispersive permittivity can be implemented by FDTD method with two steps. The first step is to update the electric flux $D$ and magnetic component $H$ within normal leap-frog procedure, and then update the $D$ and electric component $E$ by turning the angular frequency $j\omega$ into time difference $\frac{\partial}{\partial t}$, following detailed procedures in ref. 9.

The near-field of the Au particle on an Au plate is calculated using FDTD method. A uniaxial anisotropic perfectly matched layer (UPML) absorbing boundary condition [14] is applied for truncation of the FDTD lattices. The above two-step update arrangement can be well adapted into the update procedures for UPML. The structure with Au particle and the substrate in Fig. 1 is discretized in space with $ds=0.2$nm by 1000×900 grids (200nm×180nm). The plasmon spectrum is obtained by Fourier transform of the recorded time evolution of the Electric field component.

Two schemes are calculated, with the illumination by perpendicular incidence and by inclined incidence (45 degree). Two substrate cases are calculated for comparison. The 0.8nm gap between the particle and the plate is vacuum ε=1 or a dielectric layer ε=6. We understand that the dielectric material has been located experimentally to prevent the Au sphere to wet and disperse on the Au substrate, although we see no large differences in enhancement between one case and the other. We think that for the enhancement both are practically the same. How can be important 0.8nm of dielectric material when the plasmon has a spatial extension of the order of 100nm.

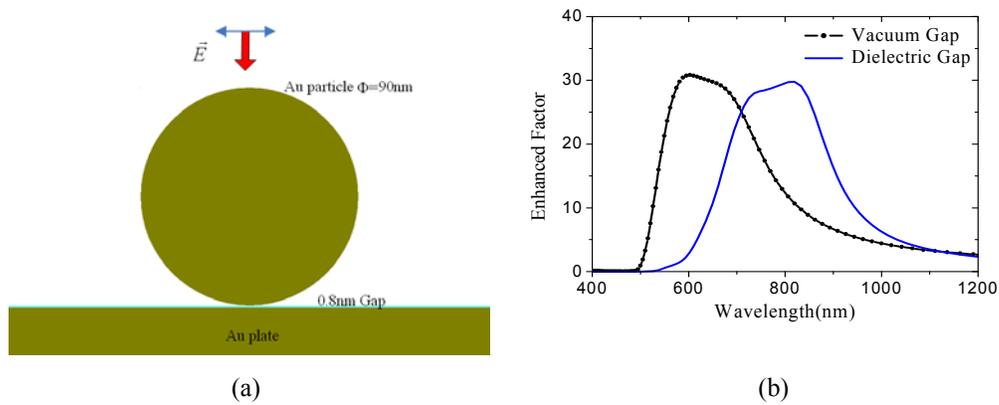

(a) (b)

FIG. 1. (a) Incident light on the nanostruture. (b) Plasmon Spectra recorded at the center of the vacuum gap, indicating a resonance at λ=600nm (dotted line); for the dielectric substrate, the resonance is at λ=830nm (solid line)

The plasmon resonance wavelengths are different for difference substrate, which are around 600nm for vacuum gap and 830nm for dielectric gap. The difference of the resonance is expected as has been studied.

To observe the field enhancement, the intensity was calculated upon the continuous plane wave incidence at the corresponding plasmon resonance wavelength. The maximum intensity does not focus in the middle of the gap, since the intensity includes the Ex component which is zero in the middle.

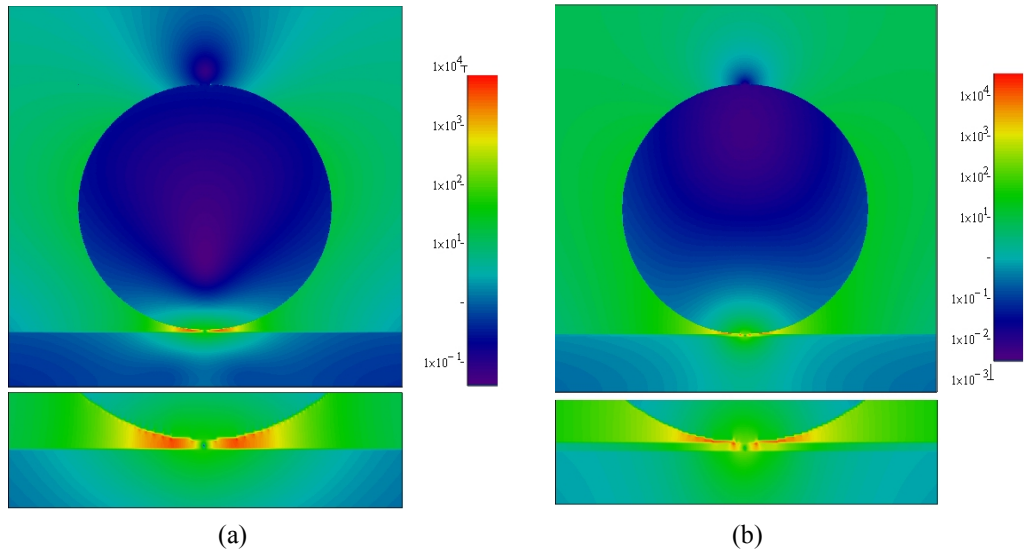

(a) (b)

FIG. 2. The intensity distribution (140nm×140nm) is mapped in terms of enhancement factor, defined as $I(x,y)/I_{inc}$. The gap area is amplified for better view. (a) Vacuum gap at λ=600nm (b)Dielectric gap at λ=830nm. The distribution is total field, with incident field included.

For inclined incidence case, as scheme shown in Fig.3(a), The plasmon resonance wavelength is retrieved about λ=850nm at center of the gap.

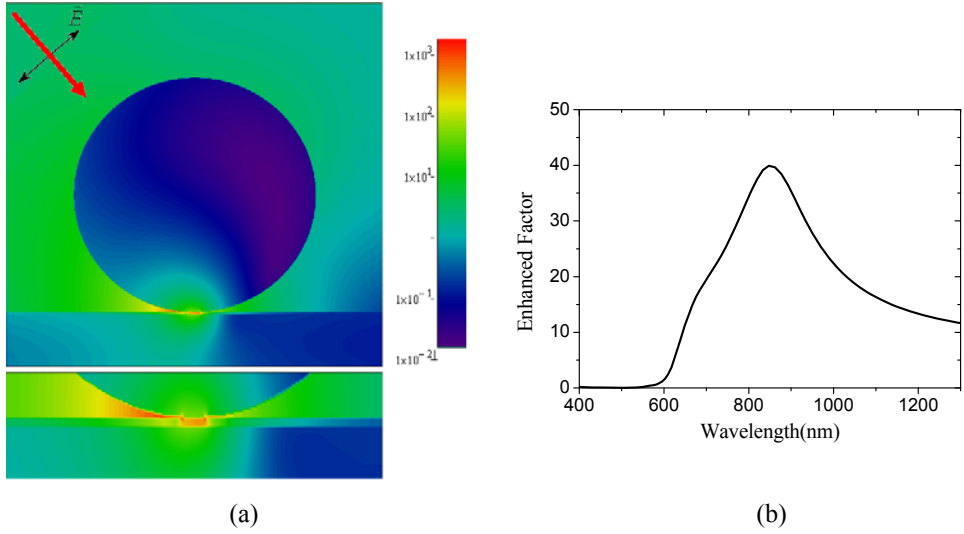

(a) (b)

FIG. 3. (a) Inclined incident scheme. Maximum intensity enhancement factor is $1.5\times10^4$. Notice that the field enhancement is the root squared of this quantity that amounts to $1.2\times10^2$. (b) The plasmon spectra at the center of the gap indicating a resonant around λ=850nm

Therefore for the case of maximum power illumination reported ($E_{laser}$=11GW/cm$^2$)[7], we obtain with the calculated enhancement approximately 0.7V/Å and not 5V/Å based on a field enhancement of $10^3$ [7], implying an intensity enhancement of $10^6$, that we certainly do not find and its effect will destruct the nanostructure and produce background as well as increase the temperature . However 0.7V/Å are the typical large values of the fields obtained for field emission [15, 16].

In order to account for the field emission, it is necessary and very important to consider the image potential. If this is not considered, the electrons always tunnel for any field because the barrier is above the energy of the electrons. However when considering the image force after a given electric field, the electrons do not tunnel and propagate without barrier creating instabilities and increasing the value of the temperature to values that remove atoms (see ref. [8,15,16]). The potential barrier with the image force plus the electric field reads,

$$V(x) = -E \cdot x - \frac{3.6}{x} \cdot Z ,  \qquad (2)$$

where the unit are in V/Å for the field $E$ (statics fields) and distance $x$. Z=1 for vacuum and Z=(ε-1)( ε+1) for dielectric materials. Plots for the barriers are presented in Fig. 4a for two values of the field 0.5V/Å and 3V/Å. It can be seen that for 0.5V/Å the barrier goes above Fermi level indicated by a horizontal line assumed to have a work function W= 4.5V. However for 3V/Å the barrier goes below the Fermi level by 2 Volts approximately. Then we have instabilities and electric breakdown for the tremendous current that flows. Imagine applying 2 Volts of potential between two points of a metal. This potential is for just one surface that here may be enough since the overlap between the surface of the sphere and the metal surface may be small enough for a separation of 0.8nm.

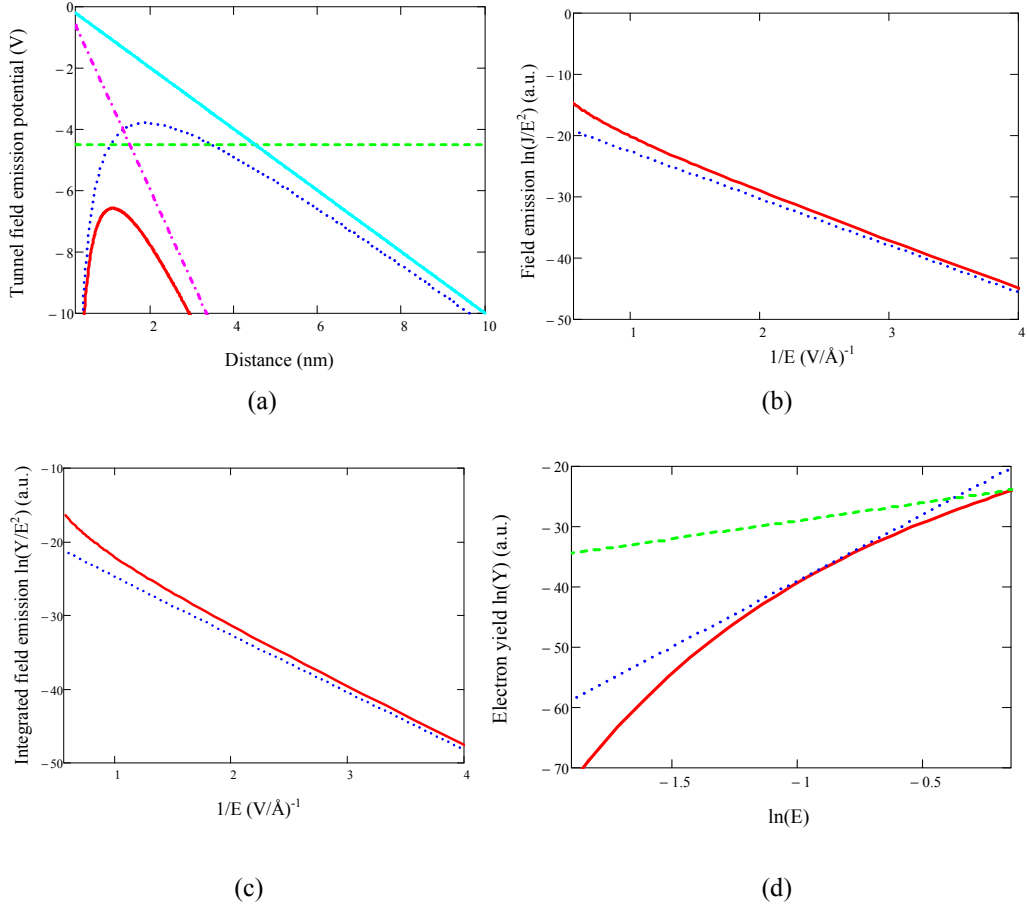

FIG. 4. (a) Tunnel field emission potential as a function of distance, with straight solid line for E=3V/Å, curved solid line for the same field with the image potential. Dashed lines are the same for 0.5V/Å. The horizontal line indicates the Fermi level of 4.5eV. (b) The field emission $\ln(J/E^2)$ as a function of $1/E$ for image potential (solid line) and without image potential (dotted line); and (c) integrated field emission $\ln(Y/E^2)$ as a function of $1/E$. (d) Double-logarithmic representation of the Y and E. The straight dashed and dotted lines indicate power laws n=1 and n=5.5, respectively. We have not introduced the experimental data because fields are off by a factor of 10 approximately.

These indicate that the field emission dominates because of the agreement between experimental data and calculations. Notice the bending up of the curves in (b) and (c) as in the experiments [7], where the potential V(x) goes above the Fermi energy (E<1.76V/Å), because of the image force.

Now we discuss the region for the values of field that are smaller than those of the fields in the current is not given by the simple the Fowler-Nordheim law. In other words this law has to be corrected as following by two functions given by elliptic integrals [6,13,14],

$$\ln(\frac{J}{E^2}) = \ln(\frac{1.55 \times 10^{10}}{Fi} \cdot t^2(w)) - \frac{0.685 \cdot W^{1.5} \cdot v(w)}{E} + Cte, \qquad (3)$$

The $t(w)$ and $v(w)$ are the elliptic functions with $w=3.8E^{0.5}/W$. The value of $t(w)$ varies between 0.8 and 1 approximately and is not important because it goes logarithmic. However $v(w)$ varies between 0, when the field produce a potential that its maximum reduces to the work function, and unity when the field is zero. Its values can be found in ref. [16].

In Fig.4b we present the results for the simple Fowler-Nordheim prediction and for the corrected case introducing the two tabulated elliptic functions as a function of the inverse of the field for W=5eV. Of course the simple function is linear but the corrected one bends up. These expressions are valid up to a field that cuts the Fermi level as discussed above. This value corresponds to E=(W/3.79)$^2$ and for W=5eV, work function of Au, corresponds to a field of 1.74V/Å. As the field is produced by the laser, it is an oscillating field of high frequency $10^{15}$/s. We have to average the field to the positive values of the amplitude to have the effective field and this is given by the integral,

$$Y = \frac{2}{T} \int_0^{T/2} J \cdot dt \qquad (4)$$

It is just up to T/2 because this is only the region that the field is positive. But this is a little confusing for *t>0* when electrons are emitted from the Au plate, but for *t<0* the electrons are emitted from the sphere in opposite direction as that for *t>0*, then the electrons are always emitted. However for all t or only for *t>0* the difference is just a factor of 2 and this goes inside a logarithm. There is also Coulomb blockade that has not been considered because some electron will go in and out of the sphere and this produce Coulomb blockade be because the small sphere will be charged and discharged. But this can be object of other study and will not be the case on the total emission to the vacuum region. We also have calculated the results of Fig.4c and the experiments seem to be in the range of field emission. As can be seen the expression with the elliptic functions bends up as happens in the experiments, this is a proof that we have field emission when theory and experiments [7] agree very well.

However for the large applied power of the laser field we have that the Keldysh [15] parameter $\gamma=\omega(2mW)^{0.5}/eE$. When $E=11GW/cm^2$ and using our calculated field enhancements

γ=0.53 approximately, the field emission is favored. However for the other two smaller powers of the laser applied in the experiments [7] ($E$=7GW/cm$^2$, $E$=3.5GW/cm$^2$) we have γ=0.67 and 0.95 and at least for the last case the photoemission should dominate. On the other hand, the formula for the criterion of Keldysh, as expressed above, cannot be right for which no analytical result exist. This will be a matter of further study and development of the Keldysh theory to find the right criterion.

There can be also ponderomotive forces [18, 19] in the experiments but these needs of further analysis. It is not clear that the ponderomotive explain all the large kinetic energies. Also looking to the amplitude distribution of fields or intensities on figures 2 and 3, we see that the fields are large on the contact areas of the sphere and the plate and smaller away of these regions. It is clear that the electron will be taken away. Although these electromagnetic fields are oscillating in time, the effect will remove the electrons from high fields. But the best thing will be having the electrons on the top of a nanostructure, not in the bottom, and one may apply an electric field to focus and control the electrons.

In this sense should continue control experiments on nanostructures an micro tips as has been already done [20-22] that combines applications of static fields with laser illuminations. one interesting point to continue these types of experiments will be to combine static fields with laser power. There are experiments [8,23], with static electric fields and have seen that, in particular for Au and for $E$>1V/Å, that there is emission of Au ions. Therefore the fields have to be smaller as we find in the calculations. Combining static and dynamic fields will be possible to have pulsed beams of emitted the electrons and ions in controllable way. More results should be presented in this direction.

**References**

[1] R. R. Freeman, P. H. Bucksbaum, H. Milchberg, S. Darack, D. Schumacher, and M. E. Geusic, Phys. Rev. Lett. **59**, 1092(1987).
[2] J. Kupersztych, P. Monchicourt, and M. Raynaud, Phys. Rev. Lett. **86**, 5180 (2001).
[3] H. Raether, *Surface Plasmons* (Springer Verlag Berlin Heidelberg New York 1988).
[4] D. L. Mills and M. Weber, Phys. Rev. B **26**, 1075 (1982).
[5] N. Garcia and B. Ming, Optics Express **10028**, 14 (2006); N. García, Optics Comm. **307**, 45, (1983); N. García, G. Diaz, J.J. Saenz and C. Ocal, Surface Sci. **342**, 143 (1983).
[6] J. C. Tsang, J. R. Kirtleyand T. N. Theis, Solid State Commun. **667**,35 (1980); H. Seki and M. R. Philpott, J. Chem. Phys. **2166**, 72 (1980).
[7] F. Schertz et al, Phys. Rev. Lett. **108**, 237602 (2012). See other references in it.
[8] Vu Thien Binh, N. Garcia, S. T. Purcell, Adv. Imag. Electron Phys. **95**, 63 (1996).
[9] A. Taflove and S. C. Hagness, *Computational Electrodynamics: The Finite-Difference Time Domain Method 2nd ed.* (Boston, MA: Artech House Publishers, 2005)
[10] Montacer Dridi and Alexandre Vial, J. Phys. D: Appl. Phys. **43**, 415102 (2010).

[11] Yizhuo Chu, Ethan Schonbrun, Tian Yang, and Kenneth B. Crozier, Appl. Phys. Lett. **93**, 181108 (2008).

[12] E. D. Palik, *Handbook of Optical Constants of Solids* (Academic, New York, 1985); P. B. Johnson and R. W. Christy, Phys. Rev. B **6**, 4370 (1972).

[13] Sanshui Xiao, Niels Asger Mortensen, Min Qiu, J. Eur. Opt. Soc., Rapid Publ. **2**, 07009 (2007).

[14] Stephen D. Gedney, IEEE Trans. on Antenn. and Propag. **44**, 1630 (1996).

[15] R. Gomer, *Field Emission and Field Ionization* (Harvard University Press Cambridge, Massachusetts, 1961).

[16] H. C. Miller, J. Franklin Inst. **282**, 382 (1966).

[17] L. Keldysh, Sov. Phys. JEPT **20**, 1307 (1965).

[18] J. Kupersztych and M. Raynaud, Phys. Rev. Lett. **95**, 147401 (2005).

[19] S. E. Irvine and A. Y. Elezzabi, Appl. Phys. Lett. **86**, 264102 (2005).

[20] M. Kruger, M. Schenk and P. Hommelhoff, NATURE (London) **475**, 78(2011).

[21] C. Ropers, D. R. Solli, C. P. Schulz, C. Lienau, and T. Elsaesser, Phys. Rev. Lett. **98**, 043907 (2007).

[22] R. Bormann et al, Phys. Rev. Lett. **105**, 147601 (2010); H- Yanagisawa et al, Phys. Rev. B **81**, 115429 (2010).

[23] Vu Thien Binh and N. Garcia, J. Microsc. I 1, 605 (1991); Ultramicroscopy **42-44**, 80 (1992) ; Vu Thien Binh et al. Phys. Rev. Lett. **69**, 2527 (1992).